Edward Simmons

As college success rates fluctuates are college students able to compete. Does this also mean that their environment is a contributing factor to their performance? In the classroom there can be many determining factors that can lead to poor, or good grades. While looking how math test achievements correlate with dropout intention, we look at higher education with statistics in a meditative view.

Statistics gives us real data to show us real results. Either in math you perform well in college with a good classroom environment. Or you have a poor performance in college with a poor classroom environment. The evidence given will help back up these standings. It will provide real results from two populations from given studies.

Based on Olena Kryshko's cross-sectional and longitude study with undergraduate students she had two populations of 249, and 210 (Kryshko, 2020). The mean of the two populations are 229.5. "According to the Motivational Regulation Model of Schwinger and Stiensmeier-Pelster (2012), we assumed that motivational regulation strategies will positively predict academic performance and negatively predict dropout intention via increased academic effort" (Kryshko, 2020). In Tosto's Bandura's triadic reciprocal causation model she used the UK population, and the age group of 16-year-olds with a population of 6689(Tosto, 216). "Intrapersonal factors were significantly associated with both test scores, even when the alternative score was taken into account. Classroom environment did not correlate with mathematics achievement once intrapersonal factors and alternative test performance were included in the model, but was associated with subject interest and academic self-concept" (Tosto, 2016). In both models with increased academic performance, achievement increased. Such things as dropout intention, and classroom environment did not correlate.

With an increase of 1% of GDP half a standard deviation increases in individual's math, and science performance (Tosto, 2016). In research it is said that that the higher, and more successful people are those who are more educated. Education creates more opportunities across the board for people around the world. This may, or may not stand well with people, but it creates wealth amongst those who do well. "In England and Wales the public examination taken at age 16 (GCSE: General Certificate of Secondary Education) really matters, having life- long implications. GCSE math's is graded from A* (A-star) to G, and Grade C is the minimum requirement for many educational and employment opportunities. Students who do not achieve a C in math's are not eligible to study certain A Level subjects (Advanced qualifications for UK 16+ year old's); will not be accepted by some technical and vocational courses; and are unlikely to be accepted at University" (Tosto, 2016). There is to be said with global figures that first year enrollment students are at a high. With the competitiveness of college, you question if a first year, or second year student can complete his, or her degree. Yet alone completing his, or her degree you would like them to get good grades, and not get on retention. Once a student gets on retention, he, or she has the risk to qualify to drop out of school. This would mean they would no longer be able to study at that given university for a given amount of time.  "However, according to the dropout statistics presented in these OECD reports, a large proportion of students are not able to complete their studies successfully. This can lead to serious negative consequences for the individuals concerned as well as for society in general Hence, much international research in different fields aims at examining a variety of contextual and individual factors that promote academic success, which comprises different aspects" (Kryshko, 2020). In previous studies student's high school GPAs were a good indicator of cognitive procedures. Further studies have shown have behavior have also became a factor. Good or bad behavior can dramatically affect a student's behavior.

     Mathematics can be a very challenging subject. Not everyone does well in it. There can be road blocks to doing well in the subject. Every individual student is different. Students have very different study methods, so each achievement outcome would not be similar. This would give varying effects in test scores especially in math, because each student would work out each problem differently. "To some extent, people differ in mathematics achievement because they vary in abilities that are important for learning mathematics. For example, individual differences in math's performance have been found to be associated with individual differences in memory processing speed intelligence, language ability and spatial skills" (Tosto, 2016). Everyone's abilities can vary on different levels. This means performance is going to go up and down, on different scales. "The pattern of past findings on the bivariate associations

between the use of different motivational regulation strategies and academic performance (mostly operationalized as high school or university grades) is less consistent. While a majority of studies reported significant positive (although rather weak) bivariate correlations between the frequency of use of some strategies and academic performance" (Kryshko, 2020). Helping students can help increase grades, but all learning behaviors are different. In correlation to math a teacher must motivate the student to do well, and learn math to be successful. "Academic self-concept reflects an individual's assessment of their own general academic self-worth, based on past performance as well as their performance relative to others Academic self-concept has been associated with both general school achievement and math's-specific achievement" (Tosto, 2016).

In Kryshko first study, she used it at a lecture. The population was at a German university of 249 students with 66% females. On average the students were 21 years of age with a standard deviation of 3.7, and in their third semester of studies with a mean of 3, and a standard deviation of 1.9 (Kryshko, 2020). The participants used self-talk to discipline themselves. "I promise myself that, after work, I will do something that I like"). Participants responded to the items according to a 5-point Likert-type format ranging from 1 (*rarely*) to 5 (*very often*). To test our hypotheses and answer the research questions, we computed the overall sum score across all eight strategies in addition to considering each specific strategy" (Kryshko, 2020). The self-talk performance, and self-consequating were the most successful. Using practical logic performance-avoiding self-talk were the least frequently used, and least successful. As a successful student talking down on ones self would be the wrong thing to do. The best motivation would be self-motivational talk. This would uplift the student to do better in their studies. "Moreover, correlational analyses largely supported the expected associations between motivational regulation, academic effort, and academic success" (Kryshko, 2020). The overall score of motivational regulation strategies correlated significantly and positively with academic effort. Academic effort correlated significantly higher with academic performance. "Although the results obtained in Study 1 on the contributions of motivational regulation strategies to university students' academic success largely supported our hypotheses, they are limited by the cross- sectional study design with all examined constructs (i.e., predictor, mediator, and outcome variables) assessed at the same point of measurement. A further limitation is our use of the self-reported current GPA for the operationalization of students' academic performance. Therefore, in Study 2, we aimed to replicate the main findings of Study 1 and to address its major limitations by means of a longitudinal study design and an objective measure of academic performance" (Kryshko, 2020). Because the findings were in support of Kryshko's support she still needed to find out the results of the dropout rates. Without the second study Kryshko's experiment would be limited without the findings of the dropout rates. Replicating the main findings of Study 1 will show how motivation, and regulation strategies help college students. If she finds her hypothesis wrong her results will be the opposite. So Kryshko choose students from a STEM discipline which has a high dropout rate. Her original sample consisted of a population of 210 first year students in civil-engineering (Kryshko, 2020). "The participants (34.8% female) were on average 20.1 years old ($SD = 2.0$). In the middle of the term, 194 students from the original sample completed another questionnaire on their use of motivational regulation strategies. Two months later, 187 participants rated their academic effort and intention to quit studies via an online questionnaire. To operationalize academic performance, we requested students' current GPA (available for 197 participants from the original sample; the others had not taken any exams yet) from the universities' examination offices at the end of the term" (Kryshko, 2020). The measures were the same as Study 1. The range for the responses were still from 1-4 and phrased like "I'm serious about dropping out". In her results the performance-approach self-talk, followed by self-consequating showed the highest means, whereas performance-avoidance self-talk and enhancement of situational interest showed the lowest means (Kryshko, 2020). Kryshko's second results, and expectations were not that far off. "As expected, the overall strategy score of motivational regulation showed significant indirect effects on university performance (Hypothesis 1) and dropout intention (Hypothesis 2) via aca- demic effort. In addition, academic effort mediated the effect of most specific strategies (except the strategy performance-avoidance self-talk) on both academic outcome variables. Beyond the significant indirect effects, we identified a significant direct effect of the strategy enhancement of personal significance on academic performance as well as significant direct effects of the strategy's performance-approach self- talk and environmental control on dropout intention" (Kryshko, 2020). Her studies showed to be successful. Academic effort, and strategies both have effects on performance and control on dropout intention. With a successful experiment this makes the findings very significant.

      Findings about inter-relationships between learning environments, achievement have been heterogenous. The aim of the study by Tosto was to increase the understanding of the relationship between math given by the GCSE, intrapersonal factors, and learning environments. "Participants were drawn from the Twins' Early Development Study (TEDS). TEDS is an on-going longitudinal study of three cohorts of twins born in 1994, 1995 and 1996. Families of twins were contacted through the Office for National Statistics and over 13,000 families were recruited across England and Wales. TEDS participants have been regularly tested throughout their lives. Data for this study was collected from 7448 twin pairs (male n = 3519) when they were 16 years old (mean = 16.48; SD = 0.27) using web-based tests and questionnaires" (Tosto, 2016). The twins, or participants were asked to answer math problems. They were given problems to solve, and were asked to choose the correct answer. This was a test of cognitive ability. Achievements through the years of the twins could be learned through these studies. After the test the participants were given questions on intrapersonal views. Classroom environment can be a huge determining factor in education. One question was asked about a television. They were not asked to solve the problem, but to rate the question on a scale of 1-4. This clearly is a measure of confidence. If a student is feeling well about his, or herself, they will most likely have the best response. "Perceived Classroom Environment assesses perceptions of classroom climate during math's lessons and was measured using 17 items drawn from PISA and. Participants were asked to think about their math's lessons in responding to items such as: "The teacher shows an interest in every student's learning"; and "There is noise and disorder". Items were rated on a 4-point scale, with low scores corresponding to negative classroom environments ($\alpha$ = 0.88, n = 2405)" (Tosto, 2016). "Intrapersonal factors were consistently more strongly associated with mathematics (average correlation r = 0.52 with both GCSE and web-tests) than Classroom Environment was. "The correlation of Classroom Environment with GCSE was slightly stronger (r =

0.28) than with web-tests (r = 0.24). Overall, Classroom Environment showed a stronger relationship with the three intrapersonal factors (average r = 0.35) than with achievement" (Tosto, 2016).

    In correlation to Hypothesis 1, and 2 in Kryshko's, and Tosto's findings, motivational regulation, and intrapersonal factors of the environment both significantly effect college performance. These findings had positive results. With different results they both had correlations though. Even though motivational regulation may be a self-interest also, the educational environment has a direct effect on how a student performs. These two go hand, and hand compared to these two findings. Educational self-interest may seem to be such a thing as "self-talk". The educational environment may seem to be a thing of how you feel at the time and moment in space. But in correlation according the studies they give positive findings to academics who succeed well.

    This was mainly to find two populations who could compete in college, or school. Using data from two studies back up this report to show the correlation between participants who did, or did not do well as students.